\documentclass[%
 reprint,
superscriptaddress,
 amsmath,amssymb,
 aps,
]{revtex4-2}

\usepackage{graphicx}
\usepackage{dcolumn}
\usepackage{bm}
\usepackage{subcaption}
\usepackage{hyperref}
\usepackage{cleveref}
\usepackage{xcolor}
\usepackage{physics}
\usepackage{amsmath}
\usepackage{xr}
\def\br{{\mathbf{r}}}
\def\bq{{\mathbf{q}}}

\newcommand{\zenodo}{\href{http://doi.org/10.5281/zenodo.18894170}{10.5281/zenodo.18894170}}

\begin{document}


\title{Machine learning the two-electron reduced density matrix in molecules and condensed phases}

\author{Jessica A. Martinez B.}
\altaffiliation{These authors contributed equally to this work.}
\affiliation{Department of Physics, Rutgers University, Newark, NJ 07102, USA}
\affiliation{Department of Chemistry, Rutgers University, Newark, NJ 07102, USA}

\author{Bhaskar Rana}
\altaffiliation{These authors contributed equally to this work.}
\affiliation{Department of Physics, Rutgers University, Newark, NJ 07102, USA}

\author{Xuecheng Shao}
\affiliation{Key Laboratory of Material Simulation Methods \& Software of Ministry of Education, College of Physics, Jilin University, Changchun 130012, China}

\author{Katarzyna Pernal}
\affiliation{Institute of Physics, Lodz University of Technology, Ul Wolczanska 217/221, Lodz 93-005, Poland}

\author{Michele Pavanello}
\email{m.pavanello@rutgers.edu}
\affiliation{Department of Physics, Rutgers University, Newark, NJ 07102, USA}

\date{\today}

\begin{abstract}
Machine learning is rapidly accelerating materials and chemical discovery, but most current models target energies, forces, or selected molecular properties rather than the underlying many-body electronic structure. Learning electronic-structure proxies, such as reduced density matrices, offers a path to surrogates that can predict a broad range of observables from a single ML model. Short of learning the full wavefunction, the two-electron reduced density matrix (2-RDM) is among the most information-rich, minimally lossy targets, providing direct access to expectation values of arbitrary one- and two-electron operators regardless of the strength of the underlying electron correlation. Here we show that learning the 2-RDM is a feasible goal, yielding exceptionally accurate models. We develop surrogates for correlated wavefunction methods (including configuration interaction and coupled cluster) that yield 2-RDMs with sufficient fidelity to provide direct, training-free access to energies and forces for driving energy-conserving molecular dynamics. To tackle realistic molecular condensed phases, we leverage a many-body expansion of the 2-RDM, using our ML models to supply the expansion terms and enabling ML-powered, coupled-cluster-quality electronic structure and energetics for large solvated systems. As a demonstration, we showcase a coupled-cluster-level electronic-structure calculation of glucose solvated by 500 water molecules achieved at Hartree-Fock cost. This work establishes a general framework for learning correlated electronic structure with high fidelity and deploying it to systems beyond the reach of conventional ab initio methods.
\end{abstract}

\maketitle

\section{Introduction}
Machine learning (ML) is rapidly reshaping atomistic simulation in chemistry and materials science. A major driver has been the emergence of accurate ML interatomic potentials and force fields \cite{behler2007generalized}, including increasingly general ``foundation'' models trained across broad chemical space \cite{batatia2025foundation,rezaee2024comparing}. At the same time, there is growing interest in ML surrogates that target {electronic-structure} objects rather than only energies and forces, for example, neural-network wavefunctions \cite{hermann_ab_2023,pfau2024accurate}, electron densities \cite{bogojeski_quantum_2020}, and reduced density matrices (RDMs) \cite{shao_machine_2023,hazra2024predicting}. Learning such objects is appealing because they can serve as {central} quantities from which many observables can be obtained, and can serve as platforms for developing transferable surrogate electronic-structure methods \cite{von_lilienfeld_retrospective_2020}.

Predictive electronic-structure calculations remain a dominant computational bottleneck in computational chemistry and materials science, especially when accounting accurately for electron correlation is required. In common implementations, Kohn-Sham density functional theory (DFT) scales roughly as $N^3$ with system size, while correlated wavefunction methods such as coupled cluster with singles and doubles (CCSD) scale as $N^6$. These costs are prohibitive for realistic systems and sampling demands, particularly in condensed phases. Approximations can reduce cost (for example, CC2 \cite{hattig2003geometry} or local-correlation variants such as DLPNO-CCSD \cite{saitow2017new}) and lower-cost electronic-structure models (e.g., extended tight binding \cite{bannwarth2021extended} or orbital-free DFT \cite{mi2023orbital}) can extend accessible length and time scales. Even so, further reductions in computational cost, together with improvements in accuracy, are essential to address the complexity of chemical processes and the breadth of chemical space required for discovering new materials and chemistry.

In this work, we pursue surrogate electronic-structure models aimed at predicting the structure of correlated electrons in molecular systems and molecular condensed phases. Our starting point is that predicting a single property (e.g., energy, dipole moment) is inherently limiting: a model trained for one observable does not automatically generalize to predict others. By contrast, the 2-RDM occupies a privileged position in many-electron theory \cite{mazziotti_two-electron_2012}. It encodes one-electron probabilities and electron-pair correlations, yields the total electronic energy by a simple contraction with one- and two-electron integrals (vide infra), and provides direct access to two-electron observables that are otherwise expensive to compute at scale, such as the electronic structure factor which is an important quantity to interpret X-ray diffraction data \cite{northey_ab_2014}. Recent work \cite{delgado-granados_machine_2025} has used variational 2-RDM calculations and Hartree--Fock as lower and upper limits to constrain a ML model for the energy. However, to our knowledge, the 2-RDM has never been the target of ML models. Conceptually, our ML objective is to learn the map connecting the external potential (equivalently, atom types and geometry) to the ground-state 2-RDM, motivated by the general framework in which ground-state observables are functionals of the external potential, since the latter determines the electronic Hamiltonian.

Two obstacles must be overcome to make 2-RDM learning practical. First, in a one-particle basis (typically Gaussian-type orbitals), the 2-RDM is a four-index tensor, and even after exploiting symmetries \cite{mazziotti_reduceddensitymatrix_2007}, the number of independent elements grows steeply with system size. Second, physical 2-RDMs must satisfy $N$-representability conditions (which are only partially known) that render variational 2-RDM methods technically demanding \cite{eugene_deprince_variational_2024}. These challenges help explain why most ML work applied to electronic structure methods has focused on density functionals, electron densities, and 1-RDMs \cite{remme2025stable,shao_machine_2023,hazra2024predicting,rana2025learning,akashi2025can}. 

Short of learning the full electronic wavefunction \cite{hermann_ab_2023,pfau2024accurate}, a machine-learned model for the 2-RDM is among the most general electronic-structure surrogates one can aim for in chemistry and materials science. Unlike the electron density or the 1-RDM, the 2-RDM provides direct access to expectation values of arbitrary one- and two-electron operators. In particular, it yields the electronic energy through simple contractions with one- and two-electron integrals, yielding energies and forces without training separate models for each observable. In this sense, accurate 2-RDM models offer a broadly transferable representation of the electronic structure with minimal information loss.

Our contributions are twofold. (i) We introduce an ML framework that directly targets the 2-RDM associated with correlated electronic-structure methods (e.g., complete active space CI (CASCI) and CCSD), with the goal of producing 2-RDMs that are extremely close to reference targets and sufficiently $N$-representable in practice to be used ``as is'' for evaluating energies, atomic forces, and two-electron operators such as structure factors \cite{northey_ab_2014}. (ii) To extend this approach to molecular condensed phases, we propose an RDM-friendly many-body expansion of the 2-RDM and combine it with ML models for the $n$-body terms, yielding accurate predictions at substantially reduced computational cost for large systems. As a demonstration of the method, we will present a calculation of a glucose molecule solvated by 500 water molecules at the CCSD level of theory.

Together, these ideas offer a practical path to surrogate models for correlated wavefunction methods that are both predictive (via a learned 2-RDM) and scalable (via the many-body expansion) enabling applications to systems where these methods are prohibitively expensive.

\begin{figure}[tp]
    \includegraphics[width=1.0\linewidth]{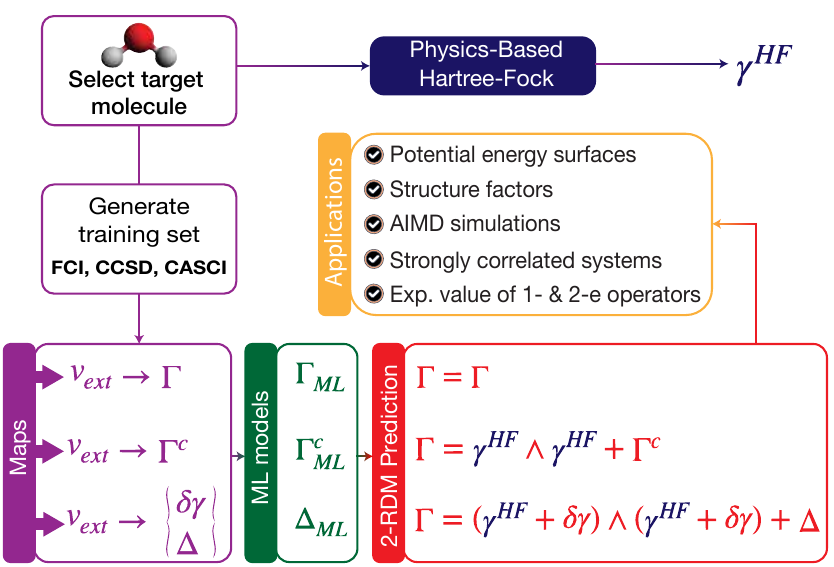}
    \caption{\label{flow} Depiction of the workflows for the $\Gamma_{\mathrm{ML}}$, $\Gamma^c_{\mathrm{ML}}$ and $\Delta_{\mathrm{ML}}$ ML models.}
\end{figure}

\section{Theory and Background}
The electronic Hamiltonian can be expressed as the sum of one- and two-electron operators expressed on an orthogonal one-electron basis set (usually HF orbitals),
\begin{equation}\label{hamiltonian}
    \hat{H} = \sum_{ps} h_{ps}\hat{a}_p^{\dagger}\hat{a}_s + \tfrac{1}{2} \sum_{psqr} g_{psqr} \hat{a}_p^{\dagger}\hat{a}_q^{\dagger}\hat{a}_r\hat{a}_s,
\end{equation}
where $h_{ps} = \langle \psi_p | -\tfrac{1}{2}\nabla^2 + v_{ext} | \psi_s \rangle$ contains the kinetic energy and the electron–nucleus attraction (external potential $v_{ext}(\br)$), and $g_{psqr}$ are the electron repulsion integrals. The operators $\hat{a}_p^{\dagger}$ and $\hat{a}_p$ denote creation and annihilation operators, respectively.

Given the ground-state $N$-electron wavefunction $\Psi_0$, the electronic energy is
\begin{equation}
E =  \bra{\Psi_0} \hat{H} \ket{\Psi_0} =  \sum_{ps} h_{ps} \gamma_{ps} + \tfrac{1}{2} \sum_{psqr} g_{psqr} \Gamma_{psqr},
\end{equation}
where the one- and two-electron reduced density matrices (1- and 2-RDMs) are defined as
\begin{equation}
    \label{rdms}
    \gamma_{ps} = \bra{\Psi_0} \hat{a}_p^{\dagger} \hat{a}_s \ket{\Psi_0}, \quad 
    \Gamma_{psqr} = \bra{\Psi_0} \hat{a}_p^{\dagger} \hat{a}_q^{\dagger} \hat{a}_r \hat{a}_s \ket{\Psi_0}.
\end{equation}

\begin{figure*}[tp]
    \includegraphics[width=1.0\textwidth]{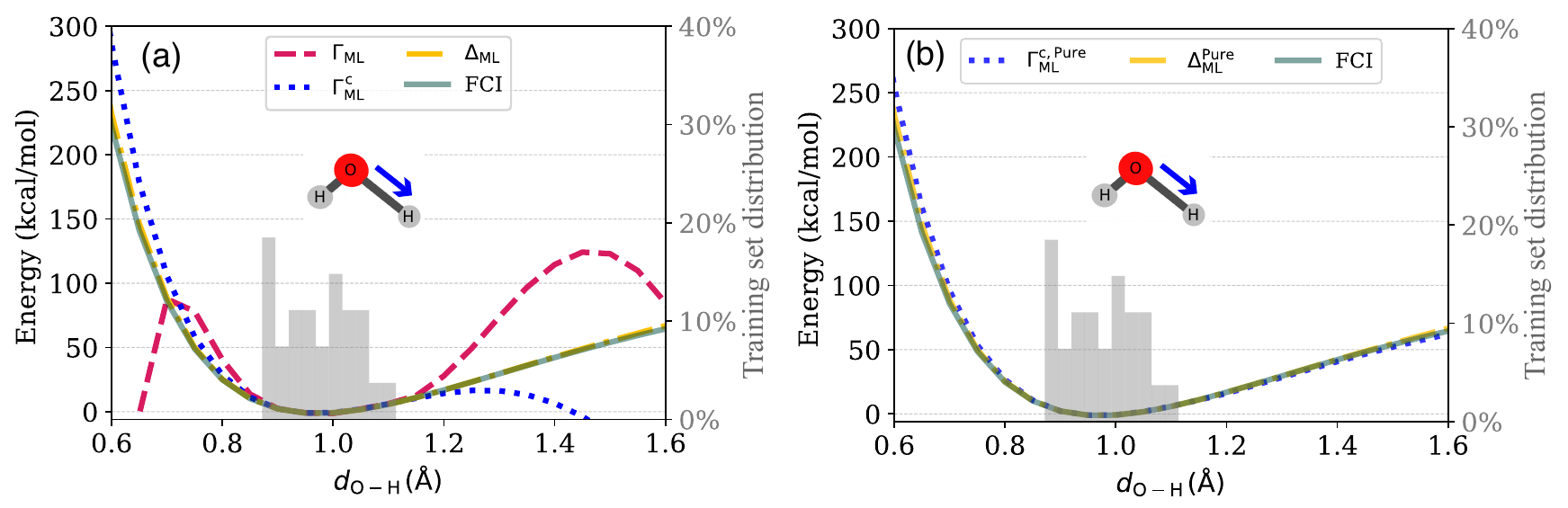}
    \caption{\label{eoh}PEC of a water molecule (FCI) as a function of the bond length of one of the OH bonds while the other is fixed at the equilibrium length. (a) Three ML models are compared. (b) the effect of purification on the 2-RDM predictions. In all panels, the distributions of training set configurations are indicated by the gray histograms. All models used the RBF kernel. For results generated with the linear kernel, see Figure \ref{sup-eoh_lin}.}
\end{figure*}

Although the energy expression above involves both the 1- and 2-RDMs, it can be recast in two equivalent forms:
\begin{align}
    \label{ene1}
    E &= {\rm Tr}\!\left[ K \Gamma \right], \quad K_{psqr} = \tfrac{1}{N-1}h_{ps}\delta_{qr} + g_{psqr},\\
    \label{ene2}
    E &= {\rm Tr}\!\left[ h \gamma \right] + {\rm Tr}\!\left[(\gamma \wedge \gamma) g \right] + {\rm Tr}\!\left[ \Delta g \right] \\
    \label{ene3}
    &= E_\mathrm{HF}[\gamma] + {\rm Tr}\!\left[ \Delta g \right].
\end{align}
where $E_{\rm HF}$ denotes the Hartree–Fock energy functional written in terms of 1-RDMs. In the spin-restricted case, the wedge product is
\begin{equation}
    (\gamma \wedge \gamma)_{pqrs} = \frac{1}{2}\left(2 \gamma_{ps}\gamma_{qr} - \gamma_{pq}\gamma_{rs}\right),
\end{equation}
where $\Delta$ is called cumulant. Thus, the 2-RDM decomposes as $\Gamma = \gamma \wedge \gamma + \Delta$, and the 1-RDM is obtained from the 2-RDM via the special trace
\begin{equation}
\gamma_{ps} = \tfrac{1}{N-1}\sum_{qr} \Gamma_{psqr}\delta_{qr} = {\rm Tr}^\prime \!\left[\Gamma\right].
\end{equation}

An important property of $\gamma$ and $\Delta$ is that they are size-extensive (see Ref.\ \citenum{kutzelnigg1999cumulant} and Supplementary Materials Section~III.A). In contrast, the non-cumulant term $\gamma \wedge \gamma$ is not.

Suppose a zeroth-order approximation to $\gamma$ is available, say $\gamma^\mathrm{HF}$ (e.g., from Hartree–Fock), such that $\gamma = \gamma^\mathrm{HF} + \delta\gamma$. The 2-RDM then becomes,
\begin{equation}
    \label{gamma}
    \Gamma = (\gamma^\mathrm{HF} + \delta \gamma ) \wedge (\gamma^\mathrm{HF} + \delta \gamma ) + \Delta.
\end{equation}
With these considerations in place, we aim to construct a machine learning model for the 2-RDM. To this end, we use the rigorous maps connecting the RDMs to the external potential $v_{ext}$ (the electron–nucleus attraction),
\begin{equation}\label{map}
    \Gamma_{\mathrm{ML}}: v_{ext} \longrightarrow \Gamma
\end{equation}
which we denote by $\Gamma_{\mathrm{ML}}$ as it will be the target of a ML model. This map follows directly from the map $v_{ext} \to \hat{H}$ by construction. Then, solving the Schr\"odinger equation gives $\hat{H} \to \Psi_0$, and from $\Psi_0$ one obtains the 2-RDM, $\Psi_0 \to \Gamma$. Thus Eq.~(\ref{map}) holds, and $\Gamma$ is a functional of the external potential, $\Gamma[v_{ext}]$. This functional need not be one-to-one, as invertibility is not required for the existence of a functional dependence.

We focus on the correlation energy, $E_c = E - E_{\rm HF}$, as well as the correlated part of the 2-RDM, $\Gamma^c = \Gamma - \gamma^\mathrm{HF} \wedge \gamma^\mathrm{HF}$. Both $E_c$ and $\Gamma^c$ can be obtained through the map in Eq.~(\ref{map}), which may be applied in two complementary ways:
\begin{align}
    \label{gammac}
    \Gamma^c_{\mathrm{ML}} &: v_{ext} \longrightarrow \Gamma^c, \\
    \label{gammad}
    \Delta_{\mathrm{ML}} &: v_{ext} \longrightarrow \delta \gamma \quad \text{and} \quad v_{ext} \longrightarrow \Delta.
\end{align}

In the following, we present, validate, and showcase the ML models for the 2-RDM for such applications as computing the molecular inelastic structure factor (including vibrational effects) of small molecules. Then, we generalize the surrogate ML models to approach condensed phases via a 2-RDM many-body expansion.

\section{ML models}

Similarly to our previous work \cite{shao_machine_2023,rana2025learning} and related approaches \cite{bogojeski_quantum_2020,brockherde_bypassing_2017,bai_machine_2022}, the ML models used to learn the maps in Eqs.~(\ref{map}), (\ref{gammac}) and (\ref{gammad}) are based on kernel ridge regression (KRR). We consider three ML strategies ($\Gamma_{\mathrm{ML}}$, $\Gamma^c_{\mathrm{ML}}$, and $\Delta_{\mathrm{ML}}$) which differ in the quantity learned and in whether a HF calculation is required at prediction time.

For example, for a generic target $\Gamma[v]$ we write
\begin{equation}
\label{krr}
\Gamma[v] = \sum_{k=1}^{N_{\mathrm{train}}} B_k\, K[v_k,v],
\end{equation}
where $v$ is the external potential at which the target is evaluated, $\{v_k\}$ are training external potentials, $B_k$ are KRR coefficients, and $K[v_k,v]$ is the kernel. We use either a linear kernel, $K_{\mathrm{LIN}}[v_k,v_l]={\rm Tr}[v_k v_l]$, or a radial basis function (RBF) kernel,
\begin{equation}
K_{\mathrm{RBF}}[v_k,v_l]=\exp\left(-\sigma\|v_k-v_l\|^2\right),
\end{equation}
where $\|\cdot\|$ denotes the Frobenius norm. The coefficients are obtained by solving the linear system $(\mathbb{K}+\alpha\mathbb{I})\mathbb{B}=\mathbb{G}$, where $\mathbb{G}$ contains the training targets and $\mathbb{K}$ is the kernel matrix. 

\begin{figure*}[!t]
    \captionsetup[subfigure]{labelformat=empty}
    \centering
    \begin{subfigure}{1.0\linewidth}
        \includegraphics[width=\linewidth]{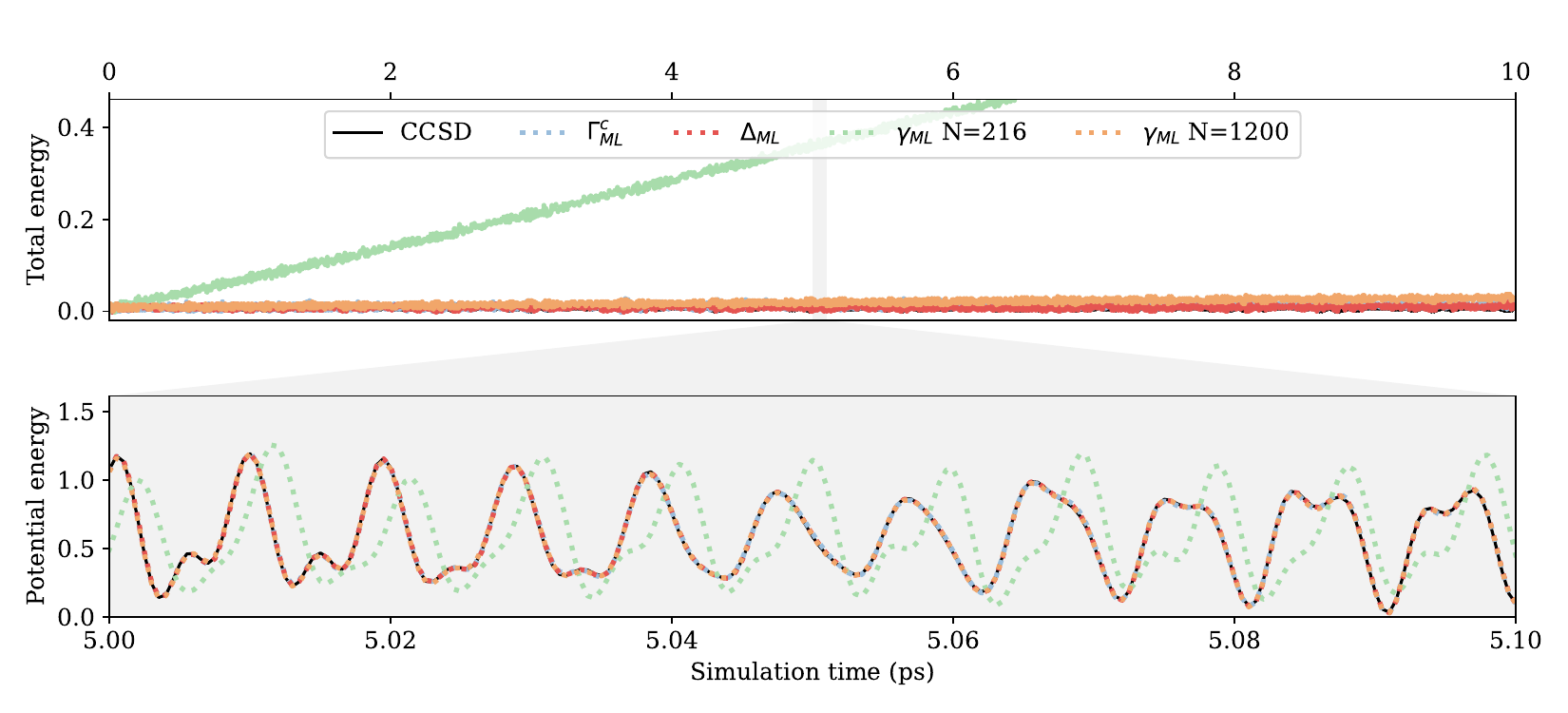}
    \end{subfigure}
    \caption{Total and potential energies in kcal/mol along an NVE trajectory for ammonia (top: total energy over 10~ps; bottom: potential energy in the interval 5.0~ps to 5.1~ps), with initial velocities sampled from a Maxwell--Boltzmann distribution at $T=300$~K. The CCSD benchmark is shown in black. The 1-RDM-based ML models, $\gamma_{{\mathrm{ML}}}$ are shown in green ($N_{\mathrm{train}}=216$) and orange ($N_{\mathrm{train}}=1200$). The $\Delta_{{\mathrm{ML}}}$ model is shown as a dotted red line and $\Gamma^{c}_{{\mathrm{ML}}}$ as a dotted blue line (both trained with $N_{\mathrm{train}}=216$ structures). Although the potential energy is not conserved in the NVE ensemble, this short time window highlights differences in its fluctuations across models. The $\gamma_{{\mathrm{ML}}}$ model with $N_{\mathrm{train}}=216$ exhibits a noticeable drift. All other models closely track the CCSD benchmark.}
    \label{ammonia_dyn}
\end{figure*}

Once trained, the three ML strategies are deployed as follows (see Figure~\ref{flow}). In $\Gamma_{\mathrm{ML}}$, a single model approximating the map in Eq.\ (\ref{map}) directly predicting the full 2-RDM at a given geometry. In $\Gamma^c_{\mathrm{ML}}$, we predict only the correlation part, $\Gamma^c$, by exploiting the map in Eq.\ (\ref{gammac}), we then perform an HF calculation at each geometry to obtain the reference 1-RDM $\gamma^\mathrm{HF}$. The full 2-RDM is then reconstructed as $\Gamma=\gamma^\mathrm{HF}\wedge\gamma^\mathrm{HF}+\Gamma^c$. In $\Delta_{\mathrm{ML}}$, we leverage the maps in Eq.\ (\ref{gammad}), and like with $\Gamma^c_{\mathrm{ML}}$, we perform an HF calculation to obtain $\gamma^\mathrm{HF}$, augment it with an ML-predicted correction $\delta\gamma$, and reconstruct the full 2-RDM from $\gamma^\mathrm{HF}$, $\delta\gamma$, and the ML-predicted $\Delta$ using Eq.~(\ref{gamma}).

Although $\Gamma_{\mathrm{ML}}$, $\Gamma^c_{\mathrm{ML}}$ and $\Delta_{\mathrm{ML}}$ are rigorous, the corresponding target quantities differ significantly in their properties, such as size extensivity and their suitability to be approximated by regression. Specifically, $\Gamma$ and $\Gamma^c$ are not size-extensive because they both contain non-cumulant terms, making them challenging learning targets. To illustrate, Table \ref{sup-tab:energy_decomp} shows the one and two-electron contributions to the total energy for the $\Delta_\mathrm{ML}$ and $\Gamma^c_\mathrm{ML}$ models clearly exposing the poorer quality regression for $\Gamma^c_\mathrm{ML}$ compared to $\Delta_\mathrm{ML}$. As argued in the Supplementary Materials Section \ref{sup-sect:se}, our models are suited for learning size extensive quantities. Thus, $\Delta_{\mathrm{ML}}$ is expected to yield the best performance, both in terms of training efficiency and model accuracy.

\section{Results for single molecules} 

\subsection{Testing ML models' $N$-representability and extrapolation behavior} We begin by considering in Figure \ref{eoh}(a) (see Figure \ref{sup-eoh_lin} for KRR carried out with a linear kernel) the potential energy curve (PEC) of a water molecule as a function of one OH bond length, keeping the other bond frozen at the equilibrium length. The ML models were trained on structures consistent with a temperature of T=300K, which, as the histogram shows in the figure, effectively clusters the geometries around the equilibrium geometry.

\begin{figure*}[!t]
    \centering
    \includegraphics[width=0.9\textwidth]{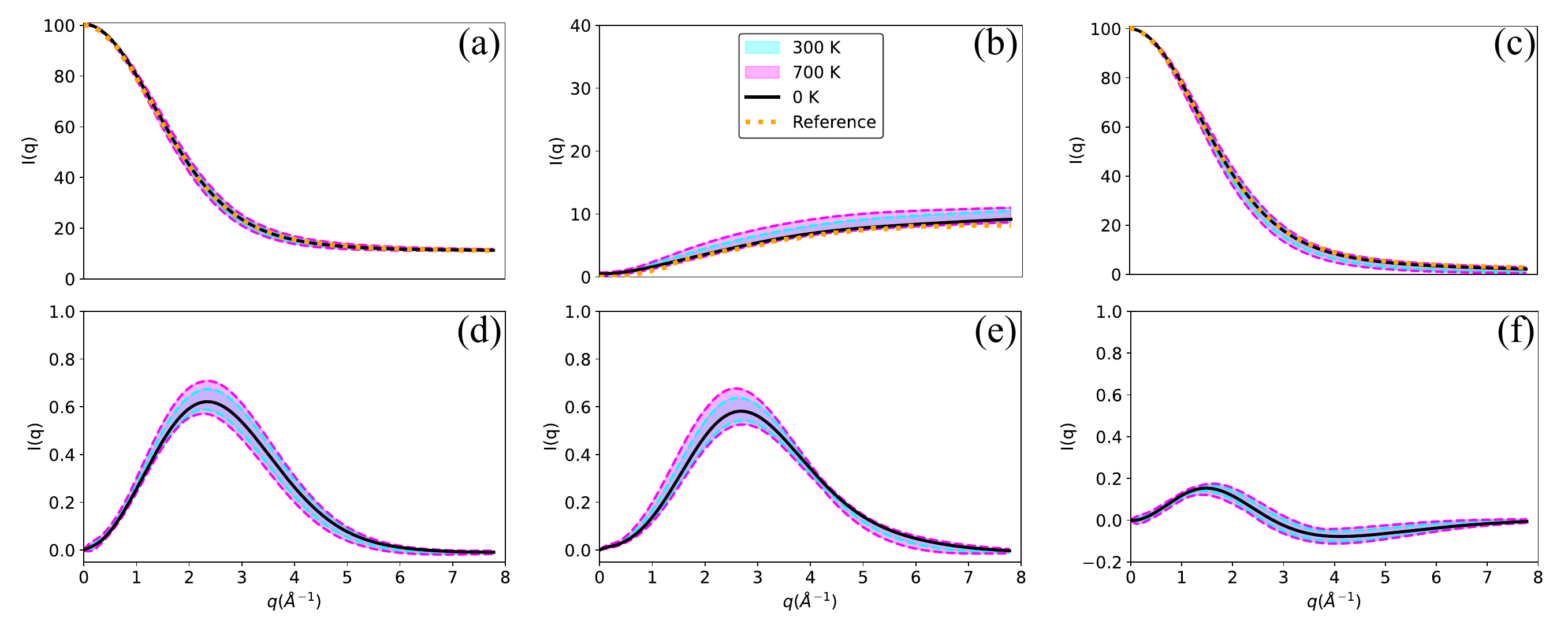}
    \caption{Effect of molecular vibrations on the electronic structure factor of gas-phase ammonia, $S_e(q)$, where $q=|\bq|$. We evaluate $S_e(q)$ using 2-RDMs predicted by the $\Delta_{\mathrm{ML}}$ model for 600 randomly sampled structures consistent with temperatures of 300 K and 700 K. Reference structure factors are taken from Ref.~\citenum{zotev_excited_2020}. Top row: (a) total $S_e(q)$ (elastic + inelastic), (b) inelastic, and (c) elastic contributions. Bottom row: corresponding contributions to $S_e(q)$ from the correlated part of the 2-RDM.}
    \label{scat_amm}
\end{figure*}

The PEC predicted by the ML models is close to the target, which is FCI, in the subspace of geometries overlapping with the training set. However, they deviate when the configurations and the training set do not overlap. Specifically, $\Gamma_{\mathrm{ML}}$ displays large deviations both for contracted and stretched OH bond lengths. The situation improves with $\Gamma^c_{\mathrm{ML}}$ where the inaccurate extrapolatory behavior mostly affects the stretched geometries. Conversely, the $\Delta_{\mathrm{ML}}$ method finds a good interpolative and extrapolative behavior.

Clearly, the $\Gamma_{\mathrm{ML}}$ and $\Gamma^c_{\mathrm{ML}}$ models are not as accurate as the $\Delta_{\mathrm{ML}}$ model. While we expect $\Gamma_{\mathrm{ML}}$ to struggle more than the other methods (as it does not rely on an Hartree-Fock result) the differences between $\Gamma^c_{\mathrm{ML}}$ and $\Delta_{\mathrm{ML}}$ are more telling of the different physical nature of the targets of these two models. As we have discussed, we expect the ML models for $\delta\gamma$ and $\Delta$ to enjoy a smooth and size extensive asymptotic behavior along bond-stretch paths. This underlying physical behavior of the target quantities results in successfully extrapolative ML models.

Purification can substantially influence the final results \cite{mazziotti_two-electron_2012,cances2008projected}. Since the data in Figure \ref{eoh}(a) were obtained without purification, we now examine Figure \ref{eoh}(b). As detailed in Supplementary Materials Sections~III.B and III.C, we purify the 2-RDM by (i) fitting the 1-RDM natural occupations to a Fermi-Dirac form and (ii) using the fitted 1-RDM to correct the 1-RDM-dependent terms in a unitary decomposition of the 2-RDM. This procedure impacts both $\Delta_{\mathrm{ML}}$ and $\Gamma^c_{\mathrm{ML}}$ models. However, we noted that $\Delta_{\mathrm{ML}}$ without purification typically provides better quality RDMs (see core energy contribution in Table \ref{sup-tab:energy_decomp}) and also benefits from error cancellation between the one- and the two-electron energy contributions in the extrapolative regime. In particular, purification restores the correct asymptotic behavior of the PEC for $\Gamma^c_{\mathrm{ML}}$ regardless of the kernel employed. Another aspect emerging from Figure \ref{eoh} is the general superiority in both interpolative and extrapolative behavior of the RBF kernel compared to the linear kernel. 

The results presented so far are specific to water. Figures {\ref{sup-co2_diss}, \ref{sup-ch3oh_diss}, \ref{sup-nh3_diss}} present PECs for CO$_2$, NH$_3$ and CH$_3$OH showing a similar behavior of the ML models considered. Hereafter, we will employ the $\Delta_{\mathrm{ML}}$ model unless otherwise stated. 

\subsection{Ab-Initio Molecular Dynamics} We tested the ability of the model RDMs to drive ab-initio molecular dynamics (AIMD) simulations. In Figure \ref{ammonia_dyn}, we present total energies (kinetic plus potential) for an isolated ammonia molecule where potential energy and forces are derived from the machine learned 2-RDMs using CCSD as our target method (see Supplementary Materials Section \ref{sup-forces_theory} for details). The stringent test consists in running separate NVE trajectories all starting from the same initial conditions. We then look for (1) the time of decoherence between the ML trajectories and the CCSD benchmark, and (2) energy drifts.

Ideally, the trajectories should be stable and coherent to the CCSD benchmark for over 10k time steps. This is when we expect the rounding errors from the finite machine precision of the CPU algebra to accumulate enough to have an effect.  Indeed, we witness coherent trajectories for the $\Delta_{\mathrm{ML}}$ and $\Gamma^c_{\mathrm{ML}}$ models which display a strict energy conservation (no appreciable energy drift is recorded in the 10~ps time window). We note that, as before, a purification step (see Methods section) for each MD step was applied to the $\Gamma^c_{\mathrm{ML}}$ RDMs while it was not required for the $\Delta_{\mathrm{ML}}$ RDMs.

We also test the models previously developed for the 1-RDM (where the map $v_{ext}\to \gamma$ is learned and then the map $\gamma\to E[\gamma], \nabla_R E$ is also learned), which we indicate with $\gamma_{\mathrm{ML}}$ in the figure. Two $\gamma_{\mathrm{ML}}$ models are considered, one using $N_{train}=216$ (for which an energy drift clearly noticeable and assessed to be 12-13 $\mathrm{K/(DOF \cdot ps)}$, see Table \ref{sup-tab:temp_drift_nh3}) and one $N_{train}=1200$.

\begin{figure*}[htp]
    \includegraphics[width=1.0\textwidth]{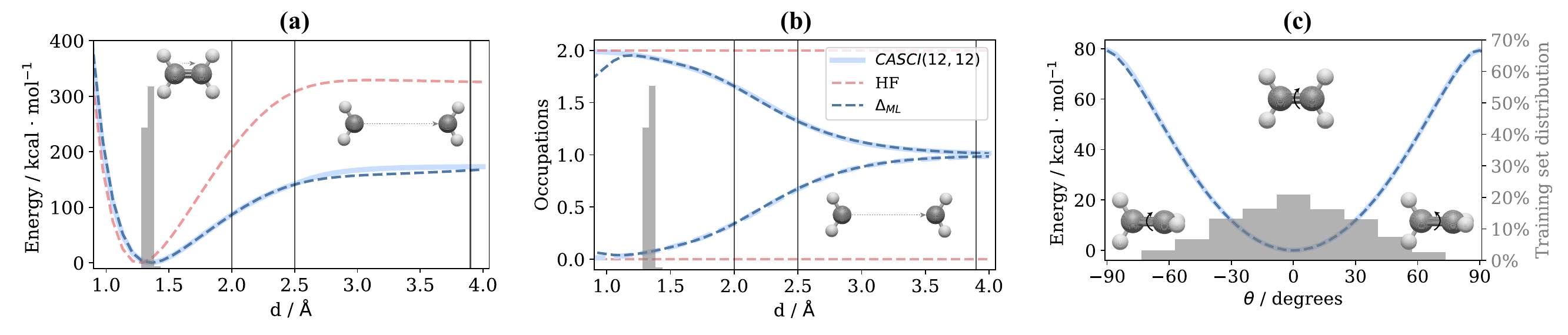} 
    \caption{\label{ethylene} Ethylene as an example of strong correlation. (a) PEC of ethylene vs C=C bond length; (b) 1-RDM occupations of HOMO ($\pi$) and LUMO ($\pi^*$) orbitals; (c) energy vs HCCH dihedral angle. In all panels, the distributions of training set configurations are indicated by the gray histograms. Vertical lines in panels (a) and (b) indicate the C=C distance of three geometries added to the training set to include in the model a notion of dissociation.}
\end{figure*}

We also run a stress test for the ML models. We trained them using training sets consistent with a temperature of T=300K and then ran AIMD trajectories with initial velocities compatible with  T=400K, 500K, and 700K, see supplementary Figures {\ref{sup-ammonia_dyn_400},  \ref{sup-ammonia_dyn_500},  \ref{sup-ammonia_dyn_700}}. The $\Delta_{\mathrm{ML}}$ and $\Gamma^c_{\mathrm{ML}}$ models once again show strong coherence with the target CCSD trajectory and stability throughout the dynamics. In contrast, the 1-RDM model for  $N_{\mathrm{train}}=216$ retains an unacceptable drift of 12 to 13 $\mathrm{K/(DOF \cdot ps)}$ across the temperature range which is addressed by increasing the number of training points to 1200. Overall, this is a strong result for the 2-RDM models. That is, even if they are trained at 300K, they deliver stable AIMDs for temperatures well beyond the training temperature requiring a reduced training set size compared to models based on the 1-RDM. Results for FCI water are consistent with this analysis and are available in Figure \ref{sup-water_dyn_700}.

\subsection{Structure Factors} 
Our results include the evaluation of the electronic structure factor, $S_{e}(\bq)$, a quantity important for understanding the structure of matter typically probed by scattering techniques, such as X-ray scattering. With the advent of X-ray free-electron lasers with pulse rate in the femtosecond timescale, it is now possible to observe real-time chemical and photochemical reactions \cite{ma2024quantitative}. This, however, is only useful if electronic structure factors can be accurately predicted for ground and excited electronic states of free molecules. As for any structure factor, they probe the presence of particles and particle pairs. The Fourier Transform of pair density, $\rho_2(\br_1,\br_2)$, is invoked and the electronic structure factor is defined as $S_{e}(\bq) = \int \rho_2(\br_1 ,\br_2) e^{i \bq \cdot (\br_2 - \br_1)}d\br_1 d\br_2 + N$. The pair density is directly available from the 2-RDM as follows $\rho_2(\br_1,\br_2) = \sum_{psqr} \Gamma_{psqr} \chi_p(\br_1)\chi_s(\br_1)\chi_q(\br_2)\chi_r(\br_2)$, where the $\chi$ functions are AOs. The evaluation of $S_e(\bq)$ is problematic as obtaining accurate pair densities or the 2-RDMs is extremely costly due to the large number of Slater determinants needed to achieve high accuracy \cite{coe_efficient_2022}. Once 2-RDMs are available, the algorithms to compute the structure factor scale like $\mathcal{O}(N^4)$ as described in the Supplementary Materials Section \ref{sup-x_ray_scat}.

As molecules are dynamical, structure factors are needed for vibrationally and electronically excited states \cite{coe_efficient_2022,zotev_excited_2020}. Accounting for the dependence of $S_e$ upon displacement from the equilibrium geometry is, in practice, a computationally even more demanding task \cite{moreno_carrascosa_ab_2019}. 
The ML models derived here, due to their excellent extrapolative behavior can efficiently sample the configuration space spanned by an AIMD simulation. In lieu of an AIMD, we generated a set of structures by generating random linear combinations of all normal modes with a probability of displacement consistent with a set temperature, using the same algorithm as the one used for generating the training set of the ML models, see the Methods section. We then proceeded to evaluate the structure factor of ammonia (see Figure \ref{scat_amm}) and water (see Figure \ref{sup-scat_wat}) in the gas phase comparing to reference values (experiment for water \cite{gallington_review_2023,amann-winkel_x-ray_2016} and a high-level ab-initio simulation for ammonia \cite{zotev_excited_2020}). Figure   \ref{sup-ammonia_widt_vs_rmse} shows that the structure factor widths are not trivially related to the RMSD of the molecular structure displacements away from the equilibrium geometry.


Figure \ref{scat_amm} shows that vibrational mode displacements cause fluctuations of up to 15\% of the intensity of the correlated part of the structure factor. Additionally, the correlated part of the inelastic structure factor shows quantitative changes upon vibrational displacement. Specifically, the peak position acquires a spread of about 0.5 {\AA}$^{-1}$, which is significant, and an equally significant spread in the peak intensity. Using ML models for predicting the 2-RDMS was crucial to produce data in a timely fashion. The computational bottleneck of the structure factor simulations in fact was the computation of the structure factor and not the generation of the correlated 2-RDMs.

\subsection{An example of strong correlation: double-bond torsion and dissociation}

When stretching covalent bonds and applying a torsion to double bonds, the onset of strong correlation is reached as soon as occupied and virtual HF orbital energies become nearly degenerate \cite{coulson1949xxxiv}. The crucial effect of such mixing is that the 1-RDM occupations drastically depart from the aufbau. In Figure \ref{ethylene}(a) we report the potential energy curve of ethylene as a function of the C-C interatomic distance, $d_{\rm CC}$. The training set used was developed as in Supplementary Materials Section \ref{sup-sampling_ethene} with the addition of three geometries at $d_{\rm CC}=$2.0, {2.5} and 3.9\AA. 

The results are quite encouraging, showing that the energy curve is well represented by the model whose training set is closely localized at the equilibrium distance (with the exception of the additional three points at dissociation). The 1-RDM occupations seen in Figure \ref{ethylene}(b) also show excellent behavior, tracing the reference non-aufbau occupations quite closely. 

Figure \ref{ethylene}(c) shows that the excellent behavior of the $\Delta_{\mathrm{ML}}$ model is retained when twisting the ethene's double bond. The figure also shows that compared with bond stretching (where we found that only a few training structures in the dissociation region were needed) the training-set distribution for bond twisting must span nearly the entire $[0,90^\circ]$ interval.

Additional examples are given in the Supplementary Materials bond stretching in ammonia, methanol, and CO$_2$, see Figures \ref{sup-nh3_diss}, \ref{sup-ch3oh_diss} and \ref{sup-co2_diss}.

\subsection{Computational Cost}
The computational cost associated with the ML models is given by the 2-RDM prediction and purification costs. The prediction cost is determined by the complexity of the KRR kernel and by the training set size. In this work, we considered linear and RBF kernels. In both cases, the cost of their evaluation is linear in the number of 2-RDM elements, which grows like $\mathcal{O}(M^4)$ with $M$ the size of the GTO basis set employed. As the RBF kernel involves the evaluation of a Gaussian function its computational complexity is slightly higher than the linear kernel. The cost associated with purification (whenever it is required) is concentrated in the diagonalization of the predicted 1-RDM. A computation that scales cubically, $\mathcal{O}(M^3)$.

\begin{figure}
    \captionsetup[subfigure]{labelformat=empty}
    \centering
    \begin{subfigure}{1.0\linewidth}
        \includegraphics[width=\linewidth]{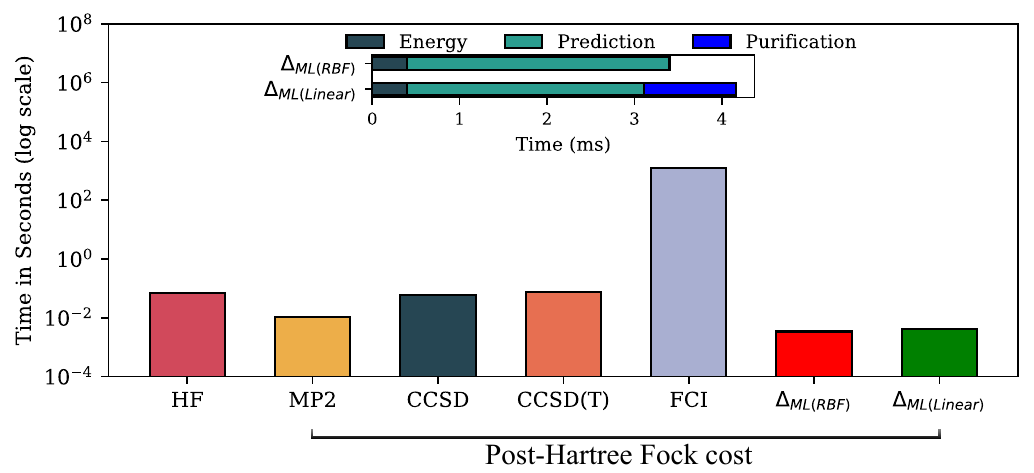}
    \end{subfigure}
    \caption{Computational cost (on a log scale) average {over 10 AIMD snapshots} of $\Delta_{\mathrm{ML}}$ for FCI 2-RDMs of water compared against the cost of HF, MP2, CCSD, CCSD(T), and FCI.  The inset shows RBF and linear kernels compared on a linear scale and split into its most important contributions: energy evaluation, prediction of the RDMs and purification (needed for the linear kernel only). All calculations are performed using a single core on the same CPU (Intel(R) Xeon(R) Platinum 8352Y CPU @ 2.20GHz).}
    \label{timings}
\end{figure}

In addition there is the unavoidable cost of contracting the RDMs with the GTO matrix elements (including the electron repulsion integrals) to obtain an electronic energy, and also the cost associated with the computation of atomic forces (see Supplementary Materials Section \ref{sup-forces_theory}). As can be seen from Figure \ref{timings} for water and Figure \ref{sup-timings_nh3} for ammonia, although the formal scaling is $\mathcal{O}(M^4)$, the computational cost is consistently lower than that of MP2.

\section{Condensed phases via RDM-based many-body expansions}
Having access to computationally cheap and accurate molecular electronic structures is a prerequisite to being able to model condensed phases. Fragmentation methods, for example, rely on ab initio solvers to compute the energy associated with each term in the $n$-body expansion of the energy. Typically, these include up to three, and even four-body energy terms \cite{herbert_fantasy_2019,paesani2026potentials}. Many-body expansions can target quantities other than the total energy, for example the correlation energy \cite{khire_pragmatic_2019,khire2023development}. Recently has emerged an approach to expand the electron density into many body terms \cite{schmittmonreal_frozendensity_2020,focke_coupled-cluster_2023}. The density is then employed in embedding workflows (using subsystem DFT). The resulting density-based many-body method, although relatively new, suggests that basing the many-body expansion on the density rather than the energy leads to faster converging energy vs many-body order of expansion. 

Borrowing this idea, enabled by the ML models presented, we consider evaluating the electronic structure and the energy of condensed phase systems by a many-body expansion involving the 2-RDM. The working equations involve the 1-RDM and the cumulant. Namely,
\begin{align}
\label{1mbrdm}
\nonumber
\gamma  &=  \gamma^\mathrm{HF} + \sum_I \delta \gamma^I + \sum_{IJ} \delta\gamma^{IJ} + \cdots \\
        &= \gamma^\mathrm{HF} + \sum_I \delta \gamma^I + \delta\gamma^{nadd} \\
\label{2mbrdm}
\Delta  &= \sum_I \Delta^I + \sum_{IJ} \Delta^{IJ} + \cdots = \sum_I \Delta^I + \Delta^{nadd}
\end{align}
where we indicated with capital letters indices spanning the fragments.

Using Eq.\ (\ref{gamma}), substituting the many-body expansions above in place of the 1-RDM and cumulant, we can recover a many-body expanded 2-RDM. 
We first consider an implementation where we truncate the expansions to one-body terms and approximate the non-additive parts of both cumulant and 1-RDM ($\Delta^{nadd}$ and $\delta\gamma^{nadd}$) with MP2. We refer to this method as MB-RDM, hereafter. We note that the one-body terms are easily obtained with our ML models exploiting the learned maps $v_I\to \{\delta\gamma^I, \Delta^I\}$ where $v_I$ is the external potential of isolated fragment $I$.

Like any method based on many-body expansions, MB-RDM is approximate and should be tested before widespread use. In Figure \ref{sup-s22e} we show MB-RDM total energies against CCSD for the 22 members of the S22 set \cite{hobza2012calculations}. Clearly, MB-RDM outperforms MP2 by one to three orders of magnitude. The RMSE of the total energies predicted by MB-RDM against CCSD is 1.3 kcal/mol, a very good result in comparison to MP2's RMSE of 33 kcal/mol.  We note that by construction, the MB-RDM errors for the interaction energies mirror the RMSE for the total energy. 

We have determined that MB-RDM successfully approximates CCSD while requiring only an MP2-like computational cost, owing to the need to evaluate the nonadditive RDMs. We now turn to solvated systems, since accurately modeling the energetics of processes in solution is known to require very large models \cite{martinez2023physical}, even when implicit solvation is employed \cite{coons2018quantum}.

\begin{figure}[!htp]
    \includegraphics[width=0.5\textwidth]{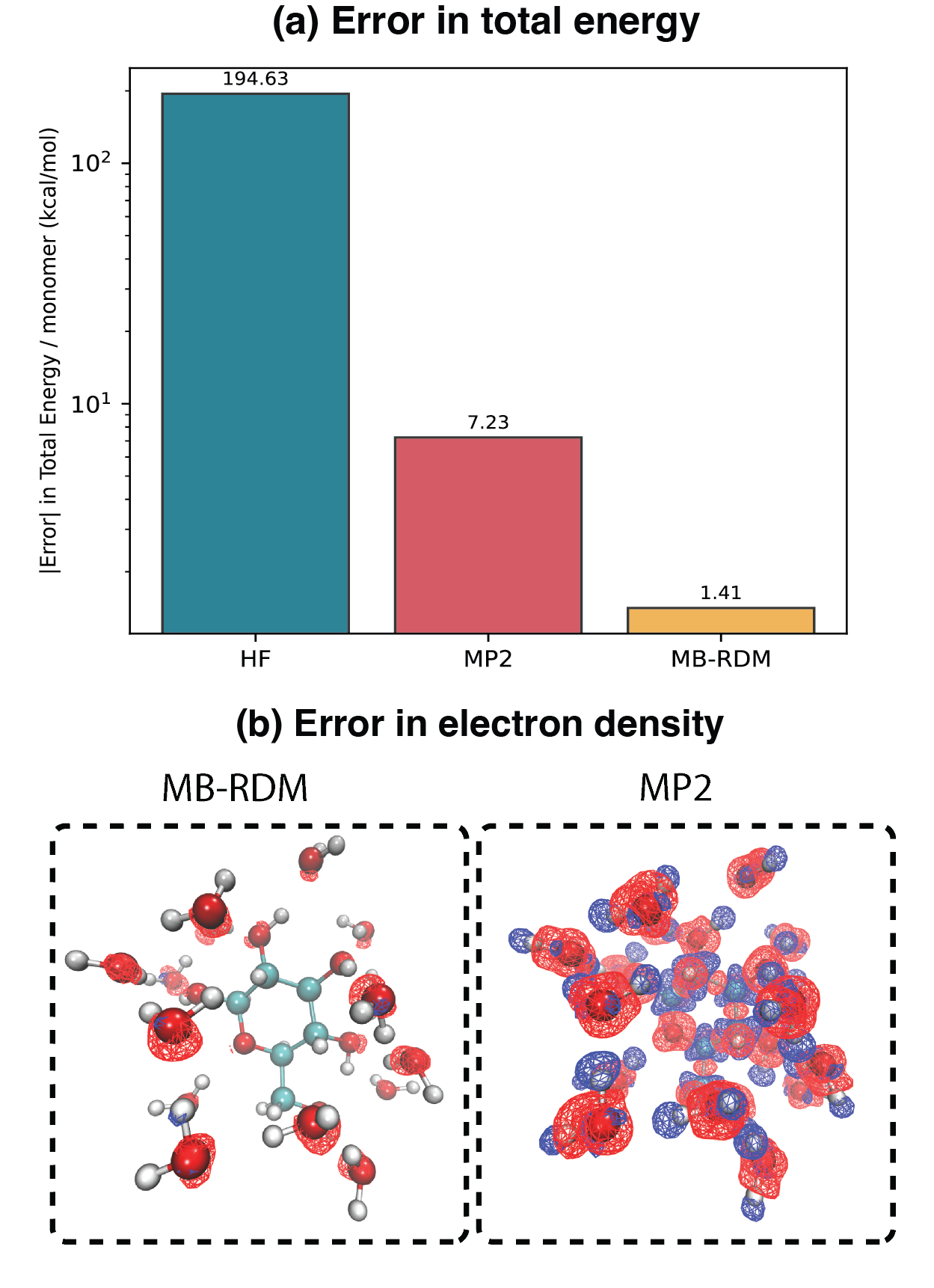}
    \caption{\label{fig:w15} Glucose solvated by 15 water molecules. Panel (a) error in the total electronic energy of HF, MP2 and MB-RDM compared against CCSD. Panel (b) electron density difference of MB-RDM and MP2 against CCSD.}
\end{figure}

Solvated systems are exactly the target of MB-RDM which can leverage ML models for the small solvent molecules such that the many-body expansion is completely delegated to ML and carried out with a very low computational cost. We first test MB-RDM on a solvated glucose molecule, (CH$_2$O)$_6$, in 15 water molecules which is a minimal model for the first solvation shell \cite{chen2025density} small enough that a full-system CCSD calculation is still feasible, and large enough that MB-RDM can be applied achieving substantial computational savings. Panel (a) of Figure \ref{fig:w15} shows that the deviation of the total electronic energy per fragment (16 fragments in total) against CCSD is 1.41 kcal/mol for MB-RDM in line with the results from the S22 set. HF and MP2, instead, deliver much worse total energies deviating by 195 and 7 kcal/mol, respectively. A similar result is apparent when comparing the electron densities differences against CCSD for MP2 and MB-RDM, see panel (b) of the figure. Clearly, the reason why MB-RDM yields more accurate energies is rooted in its more accurate electronic structure compared to MP2. 

\begin{figure*}
    \centering
    \includegraphics[width=1.0\linewidth]{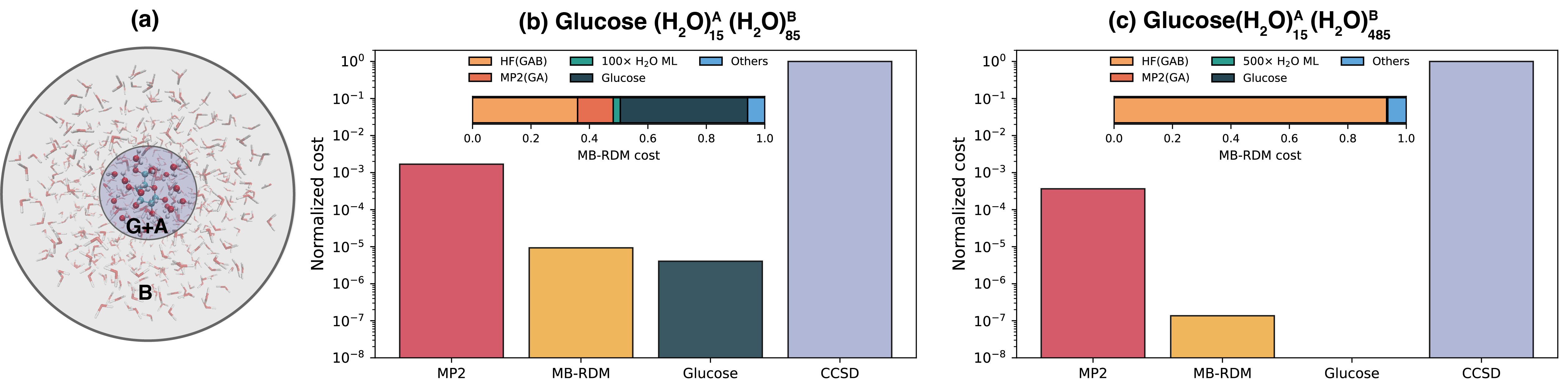}
    \caption{MB-RDM for very large systems. (a) The full system is partitioned into regions: G (glucose) and solvent regions A/B. Regions A/B define where many-body (non-additive) contributions to the cumulant and the correlated 1-RDM are treated at the MP2 level or neglected, respectively. (b,c) Computational-cost breakdown for MB-RDM for glucose solvated in water, where A is the first solvation shell (15 water molecules) and B is the remaining solvent: 85 water molecules in (b) and 485 water molecules in (c). For (b) and (c) both MP2 and CCSD costs are estimated. The ratio of the cost of glucose wrt CCSD of the full system in (c) is less than 10$^{-8}$.}
    \label{fig:w100}
\end{figure*}

A key question remains about scaling: can MB-RDM’s wall time be kept low for very large systems? The MB-RDM formulation in Eqs.~(\ref{1mbrdm}) and (\ref{2mbrdm}) requires HF and MP2 solutions for the full system (the former to compute $\gamma^{\mathrm{HF}}$ and the latter to obtain the nonadditive RDMs).
 While HF carries at most a quartic cost, MP2 is more expensive, even in its most computationally amenable formulations due to a large prefactor compared to HF \cite{mardirossian2018lowering}. Borrowing the Walter Kohn idea that electronic correlation is ``near sighted'' \cite{prodan2005nearsightedness}, we posit that the non-additive cumulant is short ranged. Evidence of this is the quantitatively small contribution of many-body dispersion in systems with gap with few exceptions \cite{ruzsinszky2012van,hermann2017first}. To approach very large systems, we modify Eqs.\ (\ref{1mbrdm}) and (\ref{2mbrdm}) as follows: the solute interacts with an inner shell of solvent molecules (region A) where non-additive (many-body) cumulant and correlated 1-RDMs are included, and an outer solvation shell (region B) where only one-body correlated 1-RDMs and cumulants are considered, see panel (a) of Figure \ref{fig:w100}.

Figure \ref{fig:w100} unsurprisingly shows that this modified MB-RDM exhibits a more favorable scaling with system size than MP2 for solvated systems; asymptotically, we expect this MB-RDM to approach HF-like scaling. For smaller clusters, such as the 101-fragment system in panel (b), the dominant MB-RDM cost arises from the monomer CCSD calculations, in particular the glucose monomer. By contrast, when region C is enlarged to 485 water molecules (panel (c)), the overall cost is dominated by the HF calculation on the full system. Replacing the monomer calculations (one-body RDMs) with ML surrogates therefore yields a substantial reduction of computational cost. Although we have not yet trained an ML model for glucose, our recent mean-field 1-RDM models already demonstrate feasibility for molecules of comparable size \cite{rana2025learning}. These results suggest that an ML-accelerated MB expansion of the RDMs for this system in its entirety, including the solute, is a realistic next step and should retain high accuracy while merely requiring mean-field cost.

\section{Conclusions}

In this work we introduced a machine-learning framework that directly targets the two-electron reduced density matrix (2-RDM) of correlated electronic-structure methods. The overarching goal is to replace expensive correlated wavefunction calculations with a surrogate ML model that preserves access to many observables, not only energies, but also forces and expectation values of one- and two-electron operators. Building on the functional dependence of ground-state observables on the external potential, we explored three complementary learning strategies that all yield the 2-RDM: direct learning of the full 2-RDM ($\Gamma_{\mathrm{ML}}$), learning of the correlated part of the 2-RDM ($\Gamma^c_{\mathrm{ML}}$), and learning of the correlated part of the 1-RDM, $\delta\gamma$, together with the cumulant, $\Delta$ ($\Delta_{\mathrm{ML}}$). Across the molecular tests considered here, the $\Delta_{\mathrm{ML}}$ strategy consistently provided the most accurate and robust predictions, reflecting the smoother and physically well-behaved asymptotic structure of its learning targets.

The $\Delta_{\mathrm{ML}}$ models closely reproduce dissociation curves for small molecules (including H$_2$O, NH$_3$, methanol, CO$_2$, and ethene), matching correlated wavefunction references even in extrapolative regimes where the tested structures lie well outside the configuration space spanned by the training set. Importantly, regimes of strong correlation (such as bond breaking and torsion of ethene's double bond) are also recovered quantitatively, both in terms of energetics and natural-orbital occupations.

Beyond static tests, we demonstrated that $\Delta_{\mathrm{ML}}$ can drive stable {ab initio} molecular dynamics trajectories with excellent energy conservation and long coherence times relative to CCSD benchmarks, even under ``stress tests'' in which trajectories are initialized at temperatures substantially above those used to generate training data. This robustness is especially encouraging for condensed-phase applications where broad sampling of configuration space is unavoidable. In addition, because the 2-RDM yields direct access to the pair-density, we leveraged the ML models to compute vibrationally broadened electronic structure factors, illustrating how learning an electronic-structure object can enable efficient, high-throughput evaluation of experimentally relevant two-electron observables that would otherwise be prohibitively expensive to obtain with traditional correlated wavefunction methods.

While the ML models presented here are already useful as drop-in surrogates for correlated RDMs, they also provide a route to shifting the computational bottleneck away from expensive solvers of correlated wavefunction methods. We demonstrated this in the context of condensed-phase chemistry via an RDM-based many-body expansion (MB-RDM), in which monomer correlated 1-RDMs and cumulants can be replaced by ML surrogates. For a range of noncovalent complexes and solvated clusters, MB-RDM yields CCSD-quality energetics and electronic structure (as probed by the electron density) when compared against full-system CCSD calculations, while retaining a substantially more favorable computational cost.

Several future directions emerge naturally from the present study. First, it will be important to reduce the formal $\mathcal{O}(M^4)$ cost associated with storing and contracting the cumulant, for example by employing density fitting/RI for the pair density and/or Cholesky-like (or other low-rank) decompositions of the 2-RDM. While conceptually straightforward, such developments may require adapting the learning targets so that the ML models directly predict the corresponding factorized representations. Second, alternative mean-field reference 1-RDMs may further improve accuracy and broaden applicability. For example, replacing $\gamma^{\mathrm{HF}}$ with a DFT reference, such as $\gamma^{\mathrm{PBE}}$ obtained with the PBE exchange-correlation functional \cite{perdew1996generalized}, is a promising direction, particularly for periodic systems.

The advances presented here provide a foundation for learning 1- and 2-RDMs and for bringing correlated wavefunction accuracy into regimes that are currently inaccessible to traditional {ab initio} solvers.

\section{Methods}

{Training sets are consistent with geometry displacements accessible at T=300K, with $N_{\mathrm{train}}=27$ for CO$_2$ and H$_2$O, $216$ for NH$_3$, and $5184$ for CH$_3$OH. We refer to Supplementary Materials Section \ref{sup-sampling_ethene}. KRR is applied using either a linear or an RBF kernel. For the linear/RBF kernel, a regularization parameter of $\alpha = 0.0045/0.0$ is used for all models. In the case of CO$_2$, the RBF kernel is employed with a length scale of $\sigma = 1.0 \times 10^{-5}$ while the others use \textsc{Scikit-learn}'s default value of $\frac{1}{M^2}$ with $M$ being the size of the AO basis set.

Whenever 2-RDM purification is considered (always required for the $\Gamma^c_{\mathrm{ML}}$ models), we first perform a unitary decomposition as in Ref.~\citenum{mazziotti_purification_2002}. We then enforce the correct normalization by replacing the trace of the 2-RDM with $N(N-1)$, and we replace every occurrence of the 1-RDM with a purified 1-RDM. The 1-RDM purification is carried out by fitting the natural occupations to a Fermi-Dirac distribution, which has been shown to yield highly accurate 1-RDMs \cite{shao_machine_2023,martinez_b_entropy_2023}. Additional details are provided in the Supplementary Materials Sections III.B and III.C.

Ab initio molecular dynamics (AIMD) simulations are performed through an API to the Atomic Simulation Environment (ASE) \cite{larsen2017atomic}. We employ the velocity Verlet integrator with a 0.5~fs time step in the microcanonical (NVE) ensemble. Initial momenta are drawn from a Maxwell-Boltzmann distribution at the target temperatures. {The 1-RDM–based machine learning models, $\gamma_{\mathrm{ML}}$, use KRR with an RBF kernel. For the map $v_{\mathrm{ext}} \rightarrow \gamma$, no regularization is applied ($\alpha = 0$). For the map $\gamma \rightarrow F$, an RBF kernel with length scale $\sigma = 1.0 \times 10^{-5}$ is employed, also with $\alpha = 0$.}

The 2-RDM learning methods developed in this work have been integrated into the all-Python QMLearn package, available on GitHub \cite{qmlearn}. QMLearn was originally developed for ML models of 1-RDMs; the present work required two key extensions: (1) additional quantum-mechanical solvers (engines) to support the workflows in Fig.~\ref{flow} based on the 2-RDM, including the evaluation of total energies and analytic energy gradients; and (2) expanded QM utilities, including unitary-decomposition-based purification and the Fourier transform of the pair density derived from the 2-RDM to compute structure factors.

QMLearn employs PySCF \cite{PYSCF} to compute all reference quantities in a Gaussian-type orbital basis set (6-31G$^*$ throughout), including external potentials, target 1- and 2-RDMs, and Hartree-Fock (HF) 1-RDMs. For H$_2$O we use Full CI (FCI), while the remaining molecules are treated with Complete Active Space Configuration Interaction (CASCI, or FOMO-CASCI for bond dissociation) using active spaces (\# orbitals, \# electrons) of (8,17) for NH$_3$, (10,16) for CO$_2$, and (12,12) for ethylene and methanol. We also report models for H$_2$O and NH$_3$ at the Coupled Cluster Singles and Doubles (CCSD) level in the same basis set. We note that to generate the reference results for the bond dissociation of ethene, we employed FOMO-CASCI \cite{slavivcek2010ab}. The active spaces for points at dissociation were generated with the maximum overlap method to be as consistent as possible with the CAS used for the training structures residing near the equilibrium geometry.\\

The data (including databases of the models in hdf5 files), jupyter notebooks, input files and raw outputs for all results presented in this paper are available at the Zenodo repository under DOI: \zenodo.

\begin{acknowledgments}
This material is based upon work supported by the US National Science Foundation under Grants No.\ CHE-2154760, OAC-1931473. Part of this research was performed while the authors were vising the Institute for Pure and Applied Mathematics (IPAM), which is supported by the National Science Foundatioon (Grant No. DMS-1925919). 
\end{acknowledgments}

\bibliography{apssamp}

@article{ma2024quantitative,
  title={Quantitative x-ray scattering of free molecules},
  author={Ma, Lingyu and Goff, Nathan and Carrascosa, Andr{\'e}s Moreno and Nelson, Silke and Liang, Mengning and Cheng, Xinxin and Yong, Haiwang and Gabalski, Ian and Huang, Lisa and Crane, Stuart W and others},
  journal={Journal of Physics B: Atomic, Molecular and Optical Physics},
  volume={57},
  number={20},
  pages={205602},
  year={2024},
  publisher={IOP Publishing}
}

@article{rana2025learning,
  title={Learning the One-Electron Reduced Density Matrix at SCF Convergence Thresholds},
  author={Rana, Bhaskar and Viot, Nicolas and Martinez B, Jessica A and Shao, Xuecheng and Ramos, Pablo and Pavanello, Michele},
  journal={Journal of Chemical Theory and Computation},
  volume={21},
  number={24},
  pages={12652--12663},
  year={2025},
  publisher={ACS Publications}
}

@article{kutzelnigg1999cumulant,
  title={Cumulant expansion of the reduced density matrices},
  author={Kutzelnigg, Werner and Mukherjee, Debashis},
  journal={The Journal of Chemical Physics},
  volume={110},
  number={6},
  pages={2800--2809},
  year={1999},
  publisher={American Institute of Physics}
}

@article{zotev_excited_2020,
	title = {Excited {Electronic} {States} in {Total} {Isotropic} {Scattering} from {Molecules}},
	volume = {16},
	issn = {1549-9618, 1549-9626},
	url = {https://pubs.acs.org/doi/10.1021/acs.jctc.9b00670},
	doi = {10.1021/acs.jctc.9b00670},
	urldate = {2023-11-07},
	journal = {Journal of Chemical Theory and Computation},
	author = {Zotev, Nikola and Moreno Carrascosa, Andrés and Simmermacher, Mats and Kirrander, Adam},
	month = apr,
	year = {2020},
	pages = {2594--2605},
}

@article{northey_ab_2014,
	title = {\textit{{Ab} {Initio}} {Calculation} of {Molecular} {Diffraction}},
	volume = {10},
	issn = {1549-9618, 1549-9626},
	url = {https://pubs.acs.org/doi/10.1021/ct500096r},
	doi = {10.1021/ct500096r},
	number = {11},
	urldate = {2023-11-07},
	journal = {Journal of Chemical Theory and Computation},
	author = {Northey, Thomas and Zotev, Nikola and Kirrander, Adam},
	month = nov,
	year = {2014},
	pages = {4911--4920},
}

@article{hermann_ab_2023,
	title = {Ab initio quantum chemistry with neural-network wavefunctions},
	volume = {7},
	issn = {2397-3358},
	url = {https://www.nature.com/articles/s41570-023-00516-8},
	doi = {10.1038/s41570-023-00516-8},
	
	number = {10},
	urldate = {2023-10-16},
	journal = {Nature Reviews Chemistry},
	author = {Hermann, Jan and Spencer, James and Choo, Kenny and Mezzacapo, Antonio and Foulkes, W. M. C. and Pfau, David and Carleo, Giuseppe and Noé, Frank},
	month = aug,
	year = {2023},
	pages = {692--709},
}

@article{bai_machine_2022,
	title = {Machine learning the {Hohenberg}-{Kohn} map for molecular excited states},
	volume = {13},
	issn = {2041-1723},
	url = {https://www.nature.com/articles/s41467-022-34436-w},
	doi = {10.1038/s41467-022-34436-w},
	
	number = {1},
	urldate = {2023-10-16},
	journal = {Nature Communications},
	author = {Bai, Yuanming and Vogt-Maranto, Leslie and Tuckerman, Mark E. and Glover, William J.},
	month = nov,
	year = {2022},
	pages = {7044},
}

@article{shao_machine_2023,
	title = {Machine learning electronic structure methods based on the one-electron reduced density matrix},
	volume = {14},
	issn = {2041-1723},
	url = {https://www.nature.com/articles/s41467-023-41953-9},
	doi = {10.1038/s41467-023-41953-9},
	
	number = {1},
	urldate = {2023-10-15},
	journal = {Nature Communications},
	author = {Shao, Xuecheng and Paetow, Lukas and Tuckerman, Mark E. and Pavanello, Michele},
	month = oct,
	year = {2023},
	pages = {6281},
}

@article{mazziotti_purification_2002,
	title = {Purification of correlated reduced density matrices},
	volume = {65},
	issn = {1063-651X, 1095-3787},
	url = {https://link.aps.org/doi/10.1103/PhysRevE.65.026704},
	doi = {10.1103/PhysRevE.65.026704},
	
	number = {2},
	urldate = {2023-09-29},
	journal = {Physical Review E},
	author = {Mazziotti, David A.},
	month = jan,
	year = {2002},
	pages = {026704},
}

@article{coe_efficient_2022,
	title = {Efficient {Computation} of {Two}-{Electron} {Reduced} {Density} {Matrices} via {Selected} {Configuration} {Interaction}},
	volume = {18},
	issn = {1549-9618, 1549-9626},
	url = {https://pubs.acs.org/doi/10.1021/acs.jctc.2c00738},
	doi = {10.1021/acs.jctc.2c00738},
	number = {11},
	urldate = {2023-05-18},
	journal = {Journal of Chemical Theory and Computation},
	author = {Coe, Jeremy P. and Moreno Carrascosa, Andrés and Simmermacher, Mats and Kirrander, Adam and Paterson, Martin J.},
	month = nov,
	year = {2022},
	pages = {6690--6699},
}

@article{mazziotti_two-electron_2012,
	title = {Two-{Electron} {Reduced} {Density} {Matrix} as the {Basic} {Variable} in {Many}-{Electron} {Quantum} {Chemistry} and {Physics}},
	volume = {112},
	issn = {0009-2665, 1520-6890},
	url = {https://pubs.acs.org/doi/10.1021/cr2000493},
	doi = {10.1021/cr2000493},
	
	number = {1},
	urldate = {2023-03-30},
	journal = {Chemical Reviews},
	author = {Mazziotti, David A.},
	month = jan,
	year = {2012},
	pages = {244--262},
}

@article{martinez_b_entropy_2023,
	title = {Entropy is a good approximation to the electronic (static) correlation energy},
	volume = {159},
	issn = {0021-9606, 1089-7690},
	url = {https://pubs.aip.org/jcp/article/159/19/191102/2921421/Entropy-is-a-good-approximation-to-the-electronic},
	doi = {10.1063/5.0171981},
	number = {19},
	urldate = {2024-01-05},
	journal = {The Journal of Chemical Physics},
	author = {Martinez B, Jessica A. and Shao, Xuecheng and Jiang, Kaili and Pavanello, Michele},
	month = nov,
	year = {2023},
	pages = {191102},
}

@article{martinez2023physical,
  title={Which Physical Phenomena Determine the Ionization Potential of Liquid Water?},
  author={Martinez B, Jessica A and Paetow, Lukas and T{\"o}lle, Johannes and Shao, Xuecheng and Ramos, Pablo and Neugebauer, Johannes and Pavanello, Michele},
  journal={The Journal of Physical Chemistry B},
  year={2023},
  volume={127},
  pages={5470},
  publisher={American Chemical Society}
}

@article{coons2018quantum,
  title={Quantum chemistry in arbitrary dielectric environments: Theory and implementation of nonequilibrium Poisson boundary conditions and application to compute vertical ionization energies at the air/water interface},
  author={Coons, Marc P and Herbert, John M},
  journal={The Journal of Chemical Physics},
  volume={148},
  number={22},
  pages={222834},
  year={2018},
  publisher={AIP Publishing}
}

@article{gallington_review_2023,
	title = {Review of {Current} {Software} for {Analyzing} {Total} {X}-ray {Scattering} {Data} from {Liquids}},
	volume = {7},
	copyright = {https://creativecommons.org/licenses/by/4.0/},
	issn = {2412-382X},
	url = {https://www.mdpi.com/2412-382X/7/2/20},
	doi = {10.3390/qubs7020020},
	number = {2},
	urldate = {2024-06-07},
	journal = {Quantum Beam Science},
	author = {Gallington, Leighanne C. and Wilke, Stephen K. and Kohara, Shinji and Benmore, Chris J.},
	month = jun,
	year = {2023},
	pages = {20},
}

@article{amann-winkel_x-ray_2016,
	title = {X-ray and {Neutron} {Scattering} of {Water}},
	volume = {116},
	issn = {0009-2665, 1520-6890},
	url = {https://pubs.acs.org/doi/10.1021/acs.chemrev.5b00663},
	doi = {10.1021/acs.chemrev.5b00663},
	
	number = {13},
	urldate = {2024-06-07},
	journal = {Chemical Reviews},
	author = {Amann-Winkel, Katrin and Bellissent-Funel, Marie-Claire and Bove, Livia E. and Loerting, Thomas and Nilsson, Anders and Paciaroni, Alessandro and Schlesinger, Daniel and Skinner, Lawrie},
	month = jul,
	year = {2016},
	pages = {7570--7589},
}

@article{bogojeski_quantum_2020,
	title = {Quantum chemical accuracy from density functional approximations via machine learning},
	volume = {11},
	issn = {2041-1723},
	url = {https://www.nature.com/articles/s41467-020-19093-1},
	doi = {10.1038/s41467-020-19093-1},
	number = {1},
	urldate = {2024-02-29},
	journal = {Nature Communications},
	author = {Bogojeski, Mihail and Vogt-Maranto, Leslie and Tuckerman, Mark E. and Müller, Klaus-Robert and Burke, Kieron},
	month = oct,
	year = {2020},
	pages = {5223},
}

@article{focke_coupled-cluster_2023,
	title = {Coupled-{Cluster} {Density}-{Based} {Many}-{Body} {Expansion}},
	volume = {127},
	copyright = {https://creativecommons.org/licenses/by/4.0/},
	issn = {1089-5639, 1520-5215},
	url = {https://pubs.acs.org/doi/10.1021/acs.jpca.3c04591},
	doi = {10.1021/acs.jpca.3c04591},
	number = {43},
	urldate = {2024-10-28},
	journal = {The Journal of Physical Chemistry A},
	author = {Focke, Kevin and Jacob, Christoph R.},
	month = nov,
	year = {2023},
	pages = {9139--9148},
}

@article{khire_pragmatic_2019,
	title = {Pragmatic {Many}-{Body} {Approach} for {Economic} {MP2} {Energy} {Estimation} of {Molecular} {Clusters}},
	volume = {123},
	copyright = {https://doi.org/10.15223/policy-029},
	issn = {1089-5639, 1520-5215},
	url = {https://pubs.acs.org/doi/10.1021/acs.jpca.9b03481},
	doi = {10.1021/acs.jpca.9b03481},
	number = {23},
	urldate = {2024-10-28},
	journal = {The Journal of Physical Chemistry A},
	author = {Khire, Subodh S. and Gadre, Shridhar R.},
	month = jun,
	year = {2019},
	pages = {5005--5011},
}

@article{khire2023development,
  title={Development and testing of an algorithm for efficient MP2/CCSD (T) energy estimation of molecular clusters with the 2--body approach},
  author={Khire, Subodh S and Gadre, Shridhar R},
  journal={Journal of Computational Chemistry},
  volume={44},
  number={3},
  pages={261--267},
  year={2023},
  publisher={Wiley Online Library}
}

@article{von_lilienfeld_retrospective_2020,
	title = {Retrospective on a decade of machine learning for chemical discovery},
	volume = {11},
	issn = {2041-1723},
	url = {https://www.nature.com/articles/s41467-020-18556-9},
	doi = {10.1038/s41467-020-18556-9},
	
	number = {1},
	urldate = {2025-01-06},
	journal = {Nature Communications},
	author = {Von Lilienfeld, O. Anatole and Burke, Kieron},
	month = sep,
	year = {2020},
	pages = {4895},
}

@article{brockherde_bypassing_2017,
	title = {Bypassing the {Kohn}-{Sham} equations with machine learning},
	volume = {8},
	copyright = {2017 The Author(s)},
	issn = {2041-1723},
	url = {https://www.nature.com/articles/s41467-017-00839-3},
	doi = {10.1038/s41467-017-00839-3},
	
	number = {1},
	urldate = {2025-01-09},
	journal = {Nature Communications},
	author = {Brockherde, Felix and Vogt, Leslie and Li, Li and Tuckerman, Mark E. and Burke, Kieron and Müller, Klaus-Robert},
	month = oct,
	year = {2017},
	pages = {872},
}

@article{eugene_deprince_variational_2024,
	title = {Variational determination of the two‐electron reduced density matrix: {A} tutorial review},
	volume = {14},
	issn = {1759-0876, 1759-0884},
	shorttitle = {Variational determination of the two‐electron reduced density matrix},
	url = {https://wires.onlinelibrary.wiley.com/doi/10.1002/wcms.1702},
	doi = {10.1002/wcms.1702},
	number = {1},
	urldate = {2025-01-09},
	journal = {WIREs Computational Molecular Science},
	author = {Eugene DePrince, A.},
	month = jan,
	year = {2024},
	pages = {e1702},
}

@article{perdew1996generalized,
  title={Generalized gradient approximation made simple},
  author={Perdew, John P and Burke, Kieron and Ernzerhof, Matthias},
  journal={Physical Review Letters},
  volume={77},
  number={18},
  pages={3865},
  year={1996},
  publisher={APS}
}

@article{cances2008projected,
  title={Projected gradient algorithms for Hartree-Fock and density matrix functional theory calculations},
  author={Canc{\`e}s, Eric and Pernal, Katarzyna},
  journal={The Journal of chemical physics},
  volume={128},
  number={13},
  pages={134108},
  year={2008},
  publisher={AIP Publishing}
}

@article{chen2025density,
  title={Density-Functionalized QM/MM Delivers Chemical Accuracy For Solvated Systems},
  author={Chen, Xin and Martinez B, Jessica A and Shao, Xuecheng and Riera Riambau, Marc and Andreussi, Oliviero and Paesani, Francesco and Pavanello, Michele},
  journal={Journal of Chemical Theory and Computation},
  year={2025},
  volume={21},
  pages={10340},
  publisher={American Chemical Society}
}

@book{mazziotti_reduceddensitymatrix_2007,
	edition = {1},
	series = {Advances in {Chemical} {Physics}},
	title = {Reduced‐{Density}‐{Matrix} {Mechanics}: {With} {Application} to {Many}‐{Electron} {Atoms} and {Molecules}},
	volume = {134},
	isbn = {9780471790563 9780470106600},
	shorttitle = {Reduced‐{Density}‐{Matrix} {Mechanics}},
	url = {https://onlinelibrary.wiley.com/doi/book/10.1002/0470106603},
	urldate = {2025-01-10},
	publisher = {Wiley},
	editor = {Mazziotti, David A.},
	month = mar,
	year = {2007},
	doi = {10.1002/0470106603},
}

@article{moreno_carrascosa_ab_2019,
	title = {Ab {Initio} {Calculation} of {Total} {X}-ray {Scattering} from {Molecules}},
	volume = {15},
	issn = {1549-9618, 1549-9626},
	url = {https://pubs.acs.org/doi/10.1021/acs.jctc.9b00056},
	doi = {10.1021/acs.jctc.9b00056},
	number = {5},
	urldate = {2023-11-07},
	journal = {Journal of Chemical Theory and Computation},
	author = {Moreno Carrascosa, Andrés and Yong, Haiwang and Crittenden, Deborah L. and Weber, Peter M. and Kirrander, Adam},
	month = may,
	year = {2019},
	pages = {2836--2846},
}

@article{herbert_fantasy_2019,
	title = {Fantasy versus reality in fragment-based quantum chemistry},
	volume = {151},
	issn = {0021-9606, 1089-7690},
	url = {https://pubs.aip.org/jcp/article/151/17/170901/198552/Fantasy-versus-reality-in-fragment-based-quantum},
	doi = {10.1063/1.5126216},
	number = {17},
	urldate = {2025-01-22},
	journal = {The Journal of Chemical Physics},
	author = {Herbert, John M.},
	month = nov,
	year = {2019},
	pages = {170901},
}

@article{schmittmonreal_frozendensity_2020,
	title = {Frozen‐density embedding‐based many‐body expansions},
	volume = {120},
	issn = {0020-7608, 1097-461X},
	url = {https://onlinelibrary.wiley.com/doi/10.1002/qua.26228},
	doi = {10.1002/qua.26228},
	number = {21},
	urldate = {2025-01-24},
	journal = {International Journal of Quantum Chemistry},
	author = {Schmitt‐Monreal, Daniel and Jacob, Christoph R.},
	month = nov,
	year = {2020},
	pages = {e26228},
}

@article{delgado-granados_machine_2025,
	title = {Machine {Learning} of {Two}-{Electron} {Reduced} {Density} {Matrices} for {Many}-{Body} {Problems}},
	copyright = {https://doi.org/10.15223/policy-029},
	issn = {1948-7185, 1948-7185},
	url = {https://pubs.acs.org/doi/10.1021/acs.jpclett.4c03366},
	doi = {10.1021/acs.jpclett.4c03366},
	
	urldate = {2025-02-27},
	journal = {The Journal of Physical Chemistry Letters},
	author = {Delgado-Granados, Luis H. and Sager-Smith, LeeAnn M. and Trifonova, Kristina and Mazziotti, David A.},
	month = feb,
	year = {2025},
	pages = {2231--2237},
}

@article{PYSCF,
	title = {{PySCF: the Python-based simulations of chemistry framework}},
	author = {Qiming Sun and Timothy C. Berkelbach and Nick S. Blunt and George H. Booth and Sheng Guo and Zhendong Li and Junzi Liu and James D. McClain and Elvira R. Sayfutyarova and Sandeep Sharma and Sebastian Wouters and Garnet Kin-Lic Chan},
	year = {2017},
	journal = {WIREs: Computational Molecular Science},
	volume = {8},
	number = {1},
	pages = {e1340}
}

@article{coulson1949xxxiv,
  title={XXXIV. Notes on the molecular orbital treatment of the hydrogen molecule},
  author={Coulson, Charles Alfred and Fischer, Inga},
  journal={The London, Edinburgh, and Dublin Philosophical Magazine and Journal of Science},
  volume={40},
  number={303},
  pages={386--393},
  year={1949},
  publisher={Taylor \& Francis}
}

@article{hobza2012calculations,
  title={Calculations on noncovalent interactions and databases of benchmark interaction energies},
  author={Hobza, Pavel},
  journal={Accounts of chemical research},
  volume={45},
  number={4},
  pages={663--672},
  year={2012},
  publisher={ACS Publications}
}

@article{larsen2017atomic,
  title={The atomic simulation environment—a Python library for working with atoms},
  author={Larsen, Ask Hjorth and Mortensen, Jens J{\o}rgen and Blomqvist, Jakob and Castelli, Ivano E and Christensen, Rune and Du{\l}ak, Marcin and Friis, Jesper and Groves, Michael N and Hammer, Bj{\o}rk and Hargus, Cory and others},
  journal={Journal of Physics: Condensed Matter},
  volume={29},
  number={27},
  pages={273002},
  year={2017},
  publisher={IOP Publishing}
}

@misc{qmlearn,
  author =        {Xuecheng Shao and Jessica Martinez and Bhaskar Rana and Michele Pavanello},
  howpublished =  {Available at https://github.com/Quantum-MultiScale/QMLearn},
  title =         {{QMLearn: A quantum machine learning electronic
                   structure method}},
  year =          {2026},
  url =           {https://github.com/Quantum-MultiScale/QMLearn},
}

@article{paesani2026potentials,
  title={From Potentials to Properties: Data-Driven Many-Body Simulations of Water and Aqueous Systems},
  author={Paesani, Francesco and Rashmi, Richa and Agnew, Henry and Zhu, Xuanyu and Bull-Vulpe, Ethan F and Zhou, Ruihan},
  journal={Annual Review of Physical Chemistry},
  volume={77},
  publisher={Annual Reviews}
}

@article{slavivcek2010ab,
  title={Ab initio floating occupation molecular orbital-complete active space configuration interaction: An efficient approximation to CASSCF},
  author={Slav{\'\i}{\v{c}}ek, Petr and Mart{\'\i}nez, Todd J},
  journal={The Journal of Chemical Physics},
  volume={132},
  number={23},
  pages={234102},
  year={2010},
  publisher={AIP Publishing}
}

@article{ruzsinszky2012van,
  title={Van der Waals coefficients for nanostructures: fullerenes defy conventional wisdom},
  author={Ruzsinszky, Adrienn and Perdew, John P and Tao, Jianmin and Csonka, G{\'a}bor I and Pitarke, Jos{\'e} Mar{\'\i}a},
  journal={Physical Review Letters},
  volume={109},
  number={23},
  pages={233203},
  year={2012},
  publisher={APS}
}

@article{prodan2005nearsightedness,
  title={Nearsightedness of electronic matter},
  author={Prodan, Emil and Kohn, Walter},
  journal={Proceedings of the National Academy of Sciences},
  volume={102},
  number={33},
  pages={11635--11638},
  year={2005},
  publisher={National Academy of Sciences}
}

@article{hermann2017first,
  title={First-principles models for van der Waals interactions in molecules and materials: Concepts, theory, and applications},
  author={Hermann, Jan and DiStasio Jr, Robert A and Tkatchenko, Alexandre},
  journal={Chemical Reviews},
  volume={117},
  number={6},
  pages={4714--4758},
  year={2017},
  publisher={ACS Publications}
}

@article{batatia2025foundation,
  title={A foundation model for atomistic materials chemistry},
  author={Batatia, Ilyes and Benner, Philipp and Chiang, Yuan and Elena, Alin M and Kov{\'a}cs, D{\'a}vid P and Riebesell, Janosh and Advincula, Xavier R and Asta, Mark and Avaylon, Matthew and Baldwin, William J and others},
  journal={The Journal of Chemical Physics},
  volume={163},
  number={18},
  pages={184110},
  year={2025},
  publisher={AIP Publishing}
}

@article{behler2007generalized,
  title={Generalized neural-network representation of high-dimensional potential-energy surfaces},
  author={Behler, J{\"o}rg and Parrinello, Michele},
  journal={Physical Review Letters},
  volume={98},
  number={14},
  pages={146401},
  year={2007},
  publisher={APS}
}

@article{rezaee2024comparing,
  title={Comparing ANI-2x, ANI-1ccx neural networks, force field, and DFT methods for predicting conformational potential energy of organic molecules},
  author={Rezaee, Mozafar and Ekrami, Saeid and Hashemianzadeh, Seyed Majid},
  journal={Scientific Reports},
  volume={14},
  number={1},
  pages={11791},
  year={2024},
  publisher={Nature Publishing Group UK London}
}

@article{hazra2024predicting,
  title={Predicting the one-particle density matrix with machine learning},
  author={Hazra, S and Patil, U and Sanvito, S},
  journal={Journal of Chemical Theory and Computation},
  volume={20},
  number={11},
  pages={4569--4578},
  year={2024},
  publisher={ACS Publications}
}

@article{hattig2003geometry,
  title={Geometry optimizations with the coupled-cluster model CC2 using the resolution-of-the-identity approximation},
  author={H{\"a}ttig, Christof},
  journal={The Journal of Chemical Physics},
  volume={118},
  number={17},
  pages={7751--7761},
  year={2003},
  publisher={American Institute of Physics}
}

@article{saitow2017new,
  title={A new near-linear scaling, efficient and accurate, open-shell domain-based local pair natural orbital coupled cluster singles and doubles theory},
  author={Saitow, Masaaki and Becker, Ute and Riplinger, Christoph and Valeev, Edward F and Neese, Frank},
  journal={The Journal of Chemical Physics},
  volume={146},
  number={16},
  pages={164105},
  year={2017},
  publisher={AIP Publishing}
}

@article{bannwarth2021extended,
  title={Extended tight-binding quantum chemistry methods},
  author={Bannwarth, Christoph and Caldeweyher, Eike and Ehlert, Sebastian and Hansen, Andreas and Pracht, Philipp and Seibert, Jakob and Spicher, Sebastian and Grimme, Stefan},
  journal={Wiley Interdisciplinary Reviews: Computational Molecular Science},
  volume={11},
  number={2},
  pages={e1493},
  year={2021},
  publisher={Wiley Online Library}
}

@article{mi2023orbital,
  title={Orbital-free density functional theory: An attractive electronic structure method for large-scale first-principles simulations},
  author={Mi, Wenhui and Luo, Kai and Trickey, SB and Pavanello, Michele},
  journal={Chemical Reviews},
  volume={123},
  number={21},
  pages={12039--12104},
  year={2023},
  publisher={American Chemical Society}
}

@article{pfau2024accurate,
  title={Accurate computation of quantum excited states with neural networks},
  author={Pfau, David and Axelrod, Simon and Sutterud, Halvard and von Glehn, Ingrid and Spencer, James S},
  journal={Science},
  volume={385},
  number={6711},
  pages={eadn0137},
  year={2024},
  publisher={American Association for the Advancement of Science}
}

@article{remme2025stable,
  title={Stable and accurate orbital-free density functional theory powered by machine learning},
  author={Remme, Roman and Kaczun, Tobias and Ebert, Tim and Gehrig, Christof A and Geng, Dominik and Gerhartz, Gerrit and Ickler, Marc K and Klockow, Manuel V and Lippmann, Peter and Schmidt, Johannes S and others},
  journal={Journal of the American Chemical Society},
  volume={147},
  number={32},
  pages={28851--28859},
  year={2025},
  publisher={ACS Publications}
}

@article{akashi2025can,
  title={Can machines learn density functionals? Past, present, and future of ML in DFT},
  author={Akashi, Ryosuke and Sogal, Mihira and Burke, Kieron},
  journal={arXiv preprint arXiv:2503.01709},
  year={2025}
}

@article{mardirossian2018lowering,
  title={Lowering of the complexity of quantum chemistry methods by choice of representation},
  author={Mardirossian, Narbe and McClain, James D and Chan, Garnet Kin},
  journal={The Journal of Chemical Physics},
  volume={148},
  number={4},
  pages={044106},
  year={2018},
  publisher={AIP Publishing}
}

\end{document}